\documentclass[12pt]{article}
\usepackage{amsmath,amsfonts}
\usepackage{algorithmic}
\usepackage{algorithm}
\usepackage{array}
\usepackage[caption=false,font=normalsize,labelfont=sf,textfont=sf]{subfig}
\usepackage{textcomp}
\usepackage{stfloats}
\usepackage{url}
\usepackage{verbatim}
\usepackage{graphicx}
\usepackage{cite}
\begin{document}

\title{Improving Channel Estimation Through Gold Sequences}

\author{Sumita Majhi, Kaushal Shelke, Pinaki Mitra, and Ujjwal Biswas 
\thanks{Sumita Majhi and Pinaki Mitra are with Department of Computer Science and Engineering
Indian Institute of Technology Guwahati,
Guwahati, Assam, India
\{sumit176101013, pinaki\}@iitg.ac.in}
\thanks{Kaushal Shelke and Ujjwal Biswas are with Department of Computer Science and Engineering
Indian Institute of Information Technology, Bhagalpur
Bhagalpur, Bihar, India
\{kaushal.2101051cs, ubiswas.cse\}@iiitbh.ac.in}
\thanks{Manuscript received April 19, 2021; revised August 16, 2021.}}

\markboth{Journal of \LaTeX\ Class Files,~Vol.~14, No.~8, August~2021}%
{Shell \MakeLowercase{\textit{et al.}}: A Sample Article Using IEEEtran.cls for IEEE Journals}


\maketitle
\begin{abstract}
This study evaluates Non-Orthogonal Multiple Access (NOMA) systems using Gold coding and Conventional-V-BLAST (C-V-BLAST). Superimposed signals on shared subcarriers make NOMA user separation difficult, unlike MIMO. Gold sequences' orthogonal features may enhance user separation and channel estimation. A novel channel estimation approach uses fractional power allocation and partially decoded data symbols. A realistic simulation environment was created using AWGN, Rayleigh fading, and shadowing. Using pilot signals, power allocation, and data symbols, our Channel Prediction Function (CPF) surpasses pilot-based techniques.
\end{abstract}

\section{Introduction}
NOMA has emerged as a viable solution for improving spectral efficiency and ensuring fairness among users in wireless communication networks. Although NOMA has notable benefits compared to classic orthogonal multiple access (OMA) schemes, its performance is greatly affected by variables such as the accuracy of channel estimates and the mitigation of interference. This study examines the efficiency of a NOMA system using Gold coding, a method that enhances user isolation and provides supplementary variety. The assumption of a common pilot channel ($H_{p}$) for all users is adopted in this chapter to analyze the fundamental challenges of channel estimation in NOMA systems, where superimposed signals inherently cause pilot contamination. While real-world scenarios involve distinct channel responses ($h_{n}$) for each user, this initial abstraction demonstrates why traditional pilot-based estimation fails under NOMA’s interference-heavy regime. Dedicated pilots consume orthogonal resources (time/frequency slots), wasting resource consumption, which eventually conflicts with the massive connectivity characteristics of NOMA. The use of dedicated pilot signals exacerbates the near-far effect in NOMA systems by causing inaccurate channel estimation for distant users due to power disparity. Gold sequences mitigate user interference through their low cross-correlation properties, while the CPF leverages power allocation and partially decoded data to refine estimates dynamically. This phased approach—first exposing the problem via simplification, then resolving it with advanced techniques—aligns with methodological best practices in NOMA research. Moreover, the paper investigates the use of deep learning for channel prediction in NOMA systems. A deep learning model is trained on a dataset to properly predict channel behavior, resulting in enhanced channel estimation and overall system performance. The proposed channel estimate approach integrates received pilot signals, power allocation vectors, and data symbols, showcasing its superiority over conventional pilot-based algorithms. This discovery has the potential to be used in different fields, such as 5G and future wireless networks. To guarantee effective and dependable NOMA communication, network operators may improve resource allocation, power control, and interference management by comprehending the influence of various noise components and channel circumstances on system performance.
In NOMA networks, effective communication requires accurate channel estimates.  Traditional pilot-based channel estimate methods used in other network topologies fail in NOMA. NOMA superimposed signals are the main justification. \cite{chen2015comparing1} introduces a pilot-based channel estimate (CE) method for NOMA networks, focusing on uplinks. It recommends superimposed pilots and iterative estimations to improve performance. Pilot contamination and limited power remain major limits. CE method for NOMA-UFMC with filter precoding is presented in \cite{singh2019novel2}. Singular Value Decomposition (SVD) effectively analyzes pilot data but has limitations of using superimposed signals and pilot power limitations. Article \cite{angjo2020channel3} examines a NOMA-OFDM downlink network. CE has "comb-type pilot subcarriers". The pilot transmission uses dedicated OFDM subcarriers. It consumes data transmission resources and reduces network efficiency. GMM-based channel estimation and signal identification demonstrate potential over pilot-based methods \cite{salari2023design4}. Understanding its computational complexity and potential constraints in similar user power levels is crucial. CS frameworks for joint channel estimation and MUD are promising for grant-free NOMA systems \cite{du2018joint5}. They require further research on their computing complexity, noise sensitivity, and possible need for additional error-correcting techniques. Several studies \cite{jia2019joint6} and \cite{gao2021low7} examined CS-based CE in NOMA networks. CS-based CE outperforms LS and MMSE. Pilot design, noise sensitivity, and error correction are important.  
\par
Existing data-aided or semi-data-aided methods may replace pilot-based channel estimate. These approaches use MIMO channel estimation's intrinsic spatial diversity \cite{kim2022semi8} \cite{jeon2020data9} \cite{liu2024large12}, reducing the requirement for pilot signals and boosting spectral efficiency. Data-aided methods may not be enough for NOMA systems, as several users share time and frequency resources. Data symbol extraction for channel estimate is difficult due to overlapping signals. For NOMA channel estimation, a paradigm change is needed. Data-aided methods are useful, but user scheduling, power allocation, and possibly restricted pilot signals must be considered. For accurate NOMA channel reconstruction, symbol selection must be based on power allocation and user distinction. In this article, we combined SNR and power allocation metrics. We created data symbols available at \cite{thz_net10}. We compared our work to \cite{kim2022semi8} in the NOMA network scenario, demonstrating better data rate and channel capacity. Our contributions are summarized:

\begin{itemize}
    \item Novel channel prediction function (CPF) that improves channel estimate accuracy using pilot signals, power allocation information, and partially decoded data symbols.
    \item A weighted average subcarrier selection technique that prioritizes subcarriers based on their possible influence on channel estimations.
    \item Introduces a novel application of Gold sequences to improve NOMA user separation and channel estimation, focusing on the problem of superimposed signals.
    \item Using simulation environment from Table 1, data symbols may be found in ~\cite{thz_net10}.
\end{itemize}

\section{System Model}
We consider a downlink NOMA network where a transmitter transmits signals to multiple receivers. We focus on a single-antenna per-user decoding approach and a $N$ number of users. The wireless channel between the transmitter and a receiver is modeled as a frequency-flat Rayleigh fading channel denoted by $ H = \left [h_{1},h_{2},\ldots, h_{N} \right ] \in  \mathbb{C}^{ N}$. We assume a block-fading channel where the elements of $H$ remain constant during a transmission frame. 
\par
A transmission block consists of a pilot subcarrier and data subcarriers. The transmitter sends a known pilot symbol $x_{p}$ on the pilot subcarrier during the pilot transmission. All users experience the same pilot channel $H_{p}$. The received pilot signal at time slot $t$ is:
\begin{equation}
 y_{p}(t) = H_p x_p + z_p(t)  
\end{equation}
Where $z_{p}(t)$ is the additive white Gaussian noise at the receiver. During data transmission, user data symbols $x_{n}^k(t)$ are transmitted on specific subcarriers $k$ within the block for user $n$. Vector power allocation $\phi_{n} = \left [  \phi_{n}^1, \phi_{n}^2, \ldots, \phi_{n}^K\right ]$ assigns power to each subcarrier $k$ for user $n$. The received signal for user $n$ at time slot $t$ is:

\begin{equation}
y_{n}(t) = \sum_{k = 1}^{K} \sqrt{\phi_{n}^k} H_{n}^k x_{n}^k(t) + \sum_{(i \neq n)} \sum_{k = 1}^{K} \sqrt{\phi_{i}^k} H_{i}^k x_{i}^k(t) + z(t)  
\end{equation}

where,

$H_{n}^k$: Channel gain for user $n$ on subcarrier $k$,
$x_{n}^k(t)$: Data symbol for user $n$ on subcarrier $k$ at time slot $t$,
$z(t)$: Additive white Gaussian noise. Here, we consider a system with a total of $K$ subcarriers. While a portion $p$ of these subcarriers is dedicated to pilot signals. The remaining subcarriers are dynamically assigned to user data. A block diagram of subcarrier allocation is illustrated in Fig. 1.

\begin{figure}[htbp]
    \centering
    \includegraphics[width=\columnwidth]{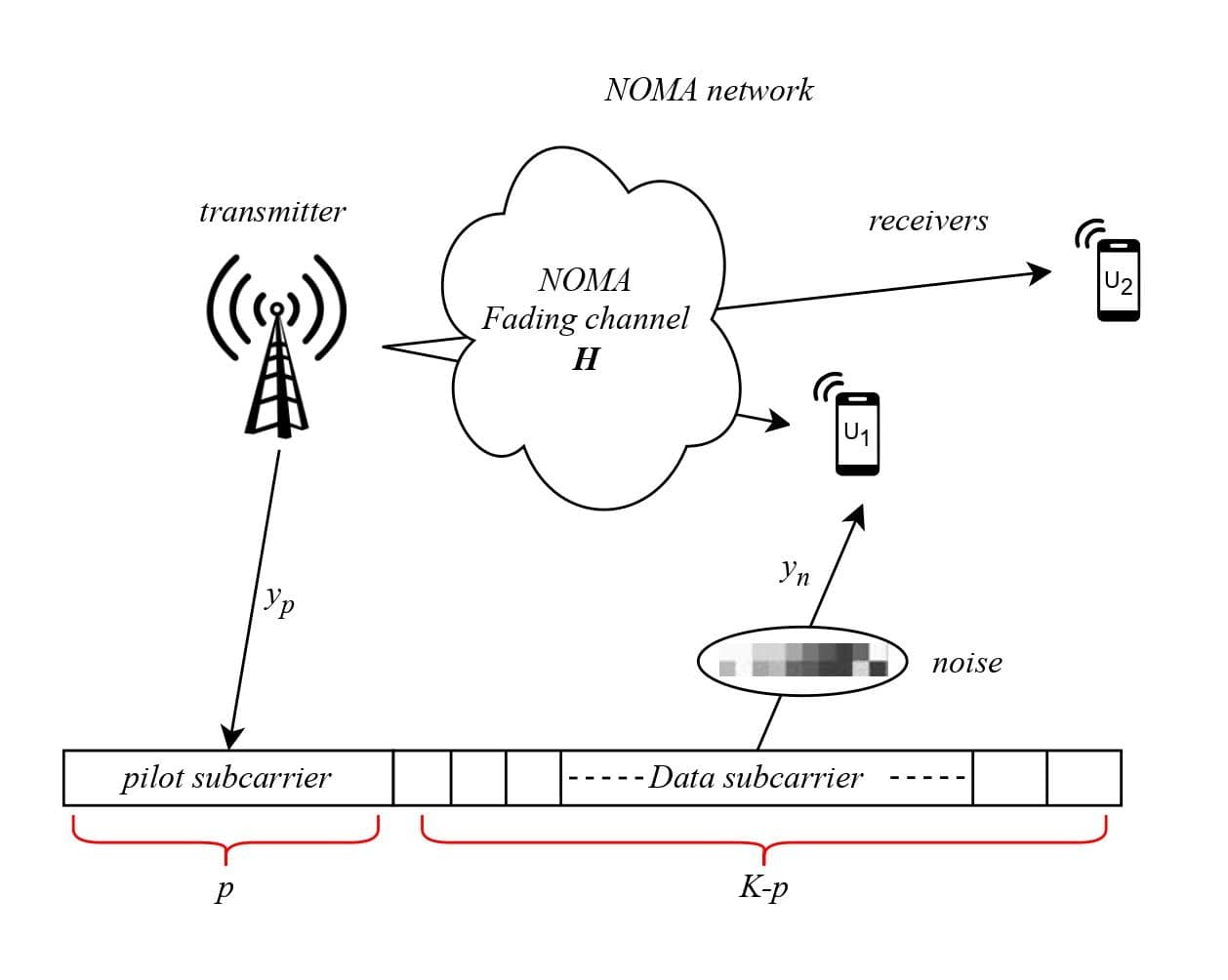} 
    \caption{Transmission block structure at the BS in NOMA network.}
\end{figure}

\par
In this work, we propose a CPF that incorporates not only the received pilot signal $y_{p}(t)$ and the allocated power vector $\phi_{n}^k(t)$, reliable partially decoded data symbols  $\hat{x}_{n}^k(t)$ for user $n$ on subcarrier $k$. Here, $\hat{x}$ denotes the estimated symbol value based on the initial decoding attempt.
\begin{equation}
H_{n}^k \text{ (predicted)} = \text{CPF}\left (\hat{h}_n^{raw}(t), \phi_{n}^k(t), \hat{x}_{n}^k(t)\right)
\end{equation}

We consider a downlink NOMA system with a single-antenna base station serving \(N\) users over frequency-flat Rayleigh fading channels (\(h_1,\ldots,h_N\)) with block fading. The transmission comprises: (1) a common pilot phase where all users share pilot \(x_p\), resulting in contaminated reception \(y_p(t) = \sum_{n=1}^N h_n x_p + z_p(t)\) (Eq.1), and (2) a data phase with superimposed user signals \(y_n(t) = \sum_{k=1}^K \sqrt{\phi_n^k} h_n x_n^k(t) + \sum_{i\neq n} \sqrt{\phi_i^k} h_i x_i^k(t) + z(t)\) (Eq.2). To address pilot contamination and interference, we propose a Gold-CP framework where: (i) each user is assigned a unique Gold code \(C_n\) (lengths \(L=31,63,127\)) for spreading (\(x = \sqrt{P}\sum_{n=1}^N S_n C_n\), Eq.4), leveraging low cross-correlation (\(|\rho_{ij}| \leq 1/\sqrt{L}\)) for initial channel estimation via despreading (\(\hat{h}_n^{raw} = y_p \star C_n^H\)); and (ii) a Channel Prediction Function (CPF) refines estimates using power allocation weights \(\phi_n^k\) (prioritizing far users via \(w_{far} > w_{near}\)) and partially decoded data \(\hat{x}_n^k\), yielding final estimates \(\hat{h}_n^{final} = CPF(\hat{h}_n^{raw}, \phi_n^k, \hat{x}_n^k)\) (Eq.3).
\section{Channel Estimation Problem in NOMA Network}
A novel strategy for downlink NOMA networks with a single pilot for two users is proposed using the well-established data-aided channel estimation method. Data blocks containing superimposing symbols from both users are utilized. The MIMO system uses a "semi-data-aided" approach \cite{kim2022semi8} to improve channel estimation accuracy. Reliable data symbols are needed for channel estimation. NOMA with solid interference may restrict the amount of properly deciphered symbols, even with advanced decoding algorithms. This unreliable data may make channel estimates difficult for users with significant interference. Instead of using all reliable symbols that may be contaminated by residual interference, we choose subcarriers with greater power allocation (fractional power) from superimposed data blocks. These subcarriers may boost transmissions. Additionally, reliable data symbols found before this stage in decoding are used. With the pilot signal, highly reliable data symbols, and high power-allocated subcarriers, channel estimation is more precise, improving NOMA system performance. 
\par
NOMA transmission frames contain superimpose symbols from multiple users, making channel estimate using only detected data symbols (even reliable ones) inadequate. Interference from the other user's superimposed signal may destroy reliable data symbols for another user. Channel estimations may be erroneous due to residual interference. However, data-aided channel estimation utilizing reliable detected symbol vectors greatly benefits MIMO systems. MIMO channels often fade independently for each path. Reliable data from one path may help estimate channel information for that path. It uses multiple antennas for spatial diversity. Data from multiple antennas might reveal channel responsiveness. Data-aided channel estimation works well in MIMO, but NOMA signals are superimposed, requiring a more advanced method.
\par
Selecting the "best" fractional power allocation for channel estimation in a NOMA network with a single pilot for two users and superimpose data blocks requires balancing reliable data for user estimation and information gain for both users. We wish to estimate channels using data symbols with a high decoding probability in the first scenario. In the second, the subcarriers should provide valuable information for estimating the close user's (good channel) and distant user's (difficult channel) channels.
\par
In Algorithm \ref{alg:channel-estimation}, we calculate a weighted average for each subcarrier. The weights of $w_{near}$ and $w_{far}$ determine the relative contribution of power allocation from each user. Sorting the weighted average ensures subcarriers with a more robust combined power allocation based on their weights appear at the beginning of the list. Finally, the algorithm selects the top $K$ subcarriers based on their weighted average for channel estimation. This approach prioritizes subcarriers with potentially stronger signals for either user while incorporating information from both users.
\begin{algorithm}
\caption{Channel Estimation using Weighted Power Allocation in NOMA}
\label{alg:channel-estimation}
\begin{algorithmic}[1]
\REQUIRE $\phi_{near}$: Vector of fractional power allocation for near user (length: number of subcarriers)\\
$\phi_{far}$: Vector of fractional power allocation for far user (length: number of subcarriers)\\
$w_{near}$: near user weighting factor $(0 \leq w_{near} \leq 1)$\\
$w_{far}$: Weighting factor for far user $(0 \leq w_{far} \leq 1)$\\
$K$: Number of subcarriers to select for channel estimation
\ENSURE Selected subcarriers: List of $K$ subcarrier indices chosen for channel estimation
\STATE Calculate Weighted Average:
\STATE Initialize an empty list $weighted_{average}$ to store the weighted average for each subcarrier.
\FOR{$k \gets 0$ to number of subcarriers $- 1$}
    \STATE Calculate the weighted average:
   using Eq (4)
\ENDFOR
\STATE Sort Subcarriers:
\STATE Sort the weighted average list in descending order. This ensures subcarriers with the highest weighted average appear first.
\STATE Select Subcarriers:
\STATE Initialize an empty list of selected subcarriers.
\FOR{$i \gets 0$ to $K - 1$}
    \STATE Add the corresponding subcarrier index from the sorted $weighted_{average}$ list to selected subcarriers.
\ENDFOR
\end{algorithmic}
\end{algorithm}
\begin{equation}
 weighted_{average}[k] = (\phi_{near}[k] \times w_{near}) + (\phi_{far}[k] \times w_{far})
\end{equation}
In our work, we adjust the weighting factors $w_{near}$ and $w_{far}$ to prioritize the far user's contribution (higher $w_{far}$ compared to $w_{near}$). This will include subcarriers with a stronger power allocation from the far user in the weighted average calculation for selecting subcarriers for channel estimation.
\begin{algorithm}
\caption{Channel Estimation using a two-step process in NOMA}
\begin{algorithmic}[1]
\REQUIRE User set $U=\{1,...,N\}$, \\
Gold code length $L_c \in \{31,63,127\}$, \\
Subcarriers $K$
\ENSURE Refined channel estimates $\hat{h}_n^{final}$

\STATE Common Pilot Transmission: 
\STATE Broadcast $x_p$: $y_p(t) = \sum_{n=1}^N h_n x_p + z_p(t)$ //Contaminated pilot

\STATE Phase 2: Gold Code Processing

\FOR{$i \gets 1$ to $N$}
     \STATE Correlate: $\hat{h}_n^{raw} = y_p * C_n^H$ //Despreading
     \STATE Apply cross-correlation bound: $|\rho_{ij}| \leq 1/\sqrt{L_c}$ 
     \STATE Weighted subcarrier ($ w_n$) selection (from Algorithm 2) 
\ENDFOR
\STATE Phase 3: Data Transmission

\STATE Allocate power: $\phi_n^k = P \cdot d_n^{-\alpha}/K$ //Fractional allocation
\STATE Transmit superimposed signal: $x = \sqrt{P}\sum_{n=1}^N S_n C_n$

\STATE Receive: $y_n(t) = \sum_{k=1}^K \sqrt{\phi_n^k} h_n x_n^k(t) + \sum_{i\neq n} \sqrt{\phi_i^k} h_i x_i^k(t) + z(t)$

\STATE Phase 4: CPF Refinement

\FOR{$i \gets 1$ to $N$}
     \STATE Perform SIC to obtain $\hat{x}_n^k$ 
     \STATE LSTM Processing: $\hat{h}_n^{final} = CPF(\hat{h}_n^{raw}, \phi_n^k, \hat{x}_n^k)$
     \STATE Update via MSE: $L = \|h_n - \hat{h}_n^{final}\|^2$
\ENDFOR
\end{algorithmic}
\end{algorithm}
\section{Gold Sequences for NOMA Channel Estimation}
NOMA networks cannot separate users using subcarrier spatial correlation like MIMO systems. When several users' data is superimposed, it's hard to tell which subcarriers reliably carry user'sdata. Using a Gold sequence on a transmission frame with one pilot and two superimpose data blocks distinguishes users' data. Gold sequences let the system recognize user data even with overlapping subcarriers. Thus NOMA network can successfully separate and decode user data. 
\par
Our work builds upon the concept of NOMA downlink channel estimation with Gold codes, where we have a single transmit antenna at the BS and a total of $N$ number of UEs with a single antenna each. Data symbol vector $S = \left (S_{1},S_{2},\ldots, S_{N}  \right )$ where $S_{n}$ represents information intended for user $n$ where $n = \left (  1,2, \ldots, N\right )$. The channel between the BS and each UE is represented by a vector $H = \left (H_{1}, H_{2},\ldots, H_{N}  \right )$, where $H_{n}$ represents the channel for user $n$. A unique code $C_{n}$ of size $L_{c}$ is assigned to user $n$ from a spreading code vector $C = \left (C_{1},C_{2},\ldots, C_{N}  \right )$. $L_{c}$ represents the Spreading code length. $P$ represents the total transmit power at the BS.
The transmitted signal $x$ is a vector of size $L_{c} \times N$ and is formed as follows:
\begin{equation}
   x = \sqrt{P} \times \sum_{n=1}^{N}S_{n} \times C_{n} 
\end{equation}
Gold codes have unique ACF and CCF features. These qualities are essential for NOMA user separation. Low out-of-phase ACF and CCF values let NOMA distinguish between different Gold-coded users' data. Similar to uplink, user separation and reliable data symbol recognition are analyzed. In this scenario, UEs can potentially identify reliable data symbols $\hat{S_{n}}$ from their received signals $y_{n}$. We may estimate each user's downlink channel at the BS using uplink channel reciprocity (assuming a quasi-static channel). Each UE transmits a pilot signal encoded with its assigned Gold code $C_{n}$ during channel estimation. The BS receives the pilot signals from all UEs and can leverage the reliable data symbols $\hat{S_{n}}$ for user $n$ to estimate the channel $(h_{n})$. We can estimate the downlink channel $\hat{h_{n}}$ of user $n$  by utilizing the received pilot signal $y_{n}^{pilot}$:
\begin{equation}
X=\left (y_{n}^{pilot} \times C_{n}^H \right )
\end{equation}
\begin{equation}
 \hat{h_{n}} = X \times diag\left (\hat{S_{n}}\right ) \times inv \left(P \times H \times C_{n} \times C_{n}^H + N \right)  
\end{equation}

where $y_{n}^{pilot}$ is the pilot signal received from user $n$. $C_{n}^H$ represents the hermitian transpose of user $n$'s spreading code.
$diag(\hat{S_{n}})$ represents the diagonal matrix with reliable data symbols of user $n$ on the diagonal. $inv()$ represents the matrix inversion and $N$ is the noise vector.
\par
Analyzing metrics like Mean Squared Error (MSE) between estimated and real channels for each user helps assess the efficiency of Gold codes. SER of each user's data after channel estimation and UE data decoding. In a NOMA downlink network with single transmit antenna, Gold codes increase user separation at UEs during decoding. This lets the BS identify reliable data symbols to estimate downlink channels for each user. This method improves channel estimates and system performance. Simulated real-world data analysis in the following section helps improve performance ratings.
\par
The initial approach from Section III of utilizing reliable data and power allocation for NOMA channel estimation may be improved iteratively using Gold sequences in Section IV. Traditional techniques have trouble distinguishing subcarrier users' data. Gold sequences' correlation qualities enable the system to differentiate users' data even when they share same subcarriers. The iterative channel estimate approach improves NOMA network performance and accuracy.
\section{Results and Discussion}
\subsection{Simulation Setup}
Table 1 delineates the Simulation Setup (left two columns) and the Model Parameters for the LSTM-based Prediction Model (right two columns).
\begin{table}[h]
\scriptsize
\caption{Simulation Setup (left two columns) and Model Parameters for LSTM-based Prediction Model (right two columns).}
\begin{tabular}{|p{.32\columnwidth}|p{.12\columnwidth}|p{.2\columnwidth}|p{.17\columnwidth}|}
\hline
\textbf{Model Parameter}                   & \textbf{Values}       & \textbf{Model Parameter}     & \textbf{Values}                \\ \hline
Number of Cells                   & 1            & Total Time Steps    & 10,000                \\ \hline
Max Transmission Power            & 43 dBm       & Phased Time Steps   & 120                   \\ \hline
Bandwidth                         & 5 MHz        & Prediction Steps    & 20                    \\ \hline
Carrier Frequency                 & 2GHz         & LSTM and GRU Layers & \{128, 64\}, 32       \\ \hline
Path Loss Exponent                & 3.76         & Dropout Rate        & 0.2, 0.3              \\ \hline
Shadowing Standard Deviation      & 10 dB        & Batch Normalization & After each layer      \\ \hline
Noise Spectral Density            & -174 dBm     & Dense Layer         & 50                    \\ \hline
Noise Figure                      & 7 dB         & Output Layer        & 20, Linear Activation \\ \hline
Min Distance BS to User           & 10 meters    & Optimize            & Adam                  \\ \hline
Number of Simulation Trials       & 10,000       & Learning rate       & 0.0008                \\ \hline
Signal-to-Noise Ratio (SNR) Range & -15 to 25 dB & Loss Function       & Mean Squared Error    \\ \hline
Distance from BS to near user     & 20           & Batch Size          & 32                    \\ \hline
Distance from BS to far user      & 50           & Epochs              & 15                    \\ \hline
\end{tabular}
\end{table}

\subsection{Performance Comparison of NOMA with Gold Coding and C-V-BLAST}
Fig. 2 shows the symbol error rate (SER) of a two-user NOMA system with single antennas using Gold sequences of lengths 31, 63, and 127. The plot illustrates the SER as a function of SNR for near and far users. Ideally, as SNR rises, both users' SER curves should fall. Due to the inherent path loss advantage of far users, SER should differ significantly between the near and far users. In difficult propagation circumstances, code length may enhance the far user's SER more than the near user's.
\begin{figure}[htbp]
    \centering
    \includegraphics[width=\columnwidth]{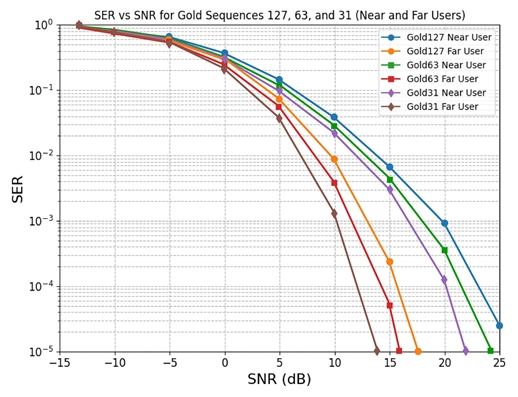} 
    \caption{SER performance comparison of a two-user NOMA system employing Gold sequence lengths 31, 63, and 127.}
\end{figure}
\par
The SER performance of a two-user NOMA system using C-V-BLAST \cite{kim2022semi8} and Gold sequences of lengths 31, 63, and 127 is compared in Fig. 3. C-V-BLAST, which uses spatial processing to mitigate interference, has limited advantages in our single-antenna system due to its restricted spatial degrees of freedom. The graph shows that gold-coded NOMA outperforms C-V-BLAST from -20 to 0 dB SNR. 
\begin{figure}[htbp]
    \centering
    \includegraphics[width=0.5\textwidth]{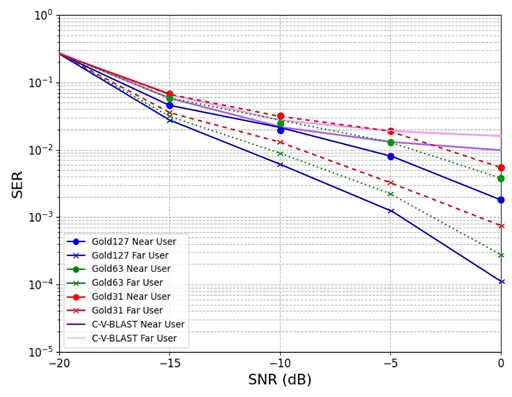} 
    \caption{SER performance comparison of Gold coding and C-V BLAST for a two-user NOMA system.}
\end{figure}

\subsection{Performance Evaluation Via Deep Learning Model}
Using latitude and longitude data from \cite{thz_net11}, we constructed a noisy dataset of 11000 data points for a realistic 5G NOMA network simulation. We simulated 5G network propagation with various noise sources.  We simulated urban multipath propagation effects with Rayleigh fading at 1.0. A log-normal shadowing component with a mean of 0.0 dB and a standard deviation of 8.0 dB was included to account for large-scale path loss. The noise model, which incorporates Rayleigh fading, and shadowing, allows us to test our NOMA system under realistic channel conditions.
\par
We captured 5G wireless channel dynamics using a 2-minute rolling window approach on the 10-minute sample in Fig. 4. This method captures 5G system-needed quick channel fluctuations while retaining adequate data for statistical analysis. Data inside this window may be used to study noise's impact on system performance, identify channel dynamics, and create 5G-specific channel estimation and decoding algorithms. The rolling window approach can study time-varying channel parameters such channel coherence time and Doppler spread, which are important 5G system design factors.
\begin{figure*}[htbp]
    \centering
    \includegraphics[width=6in]{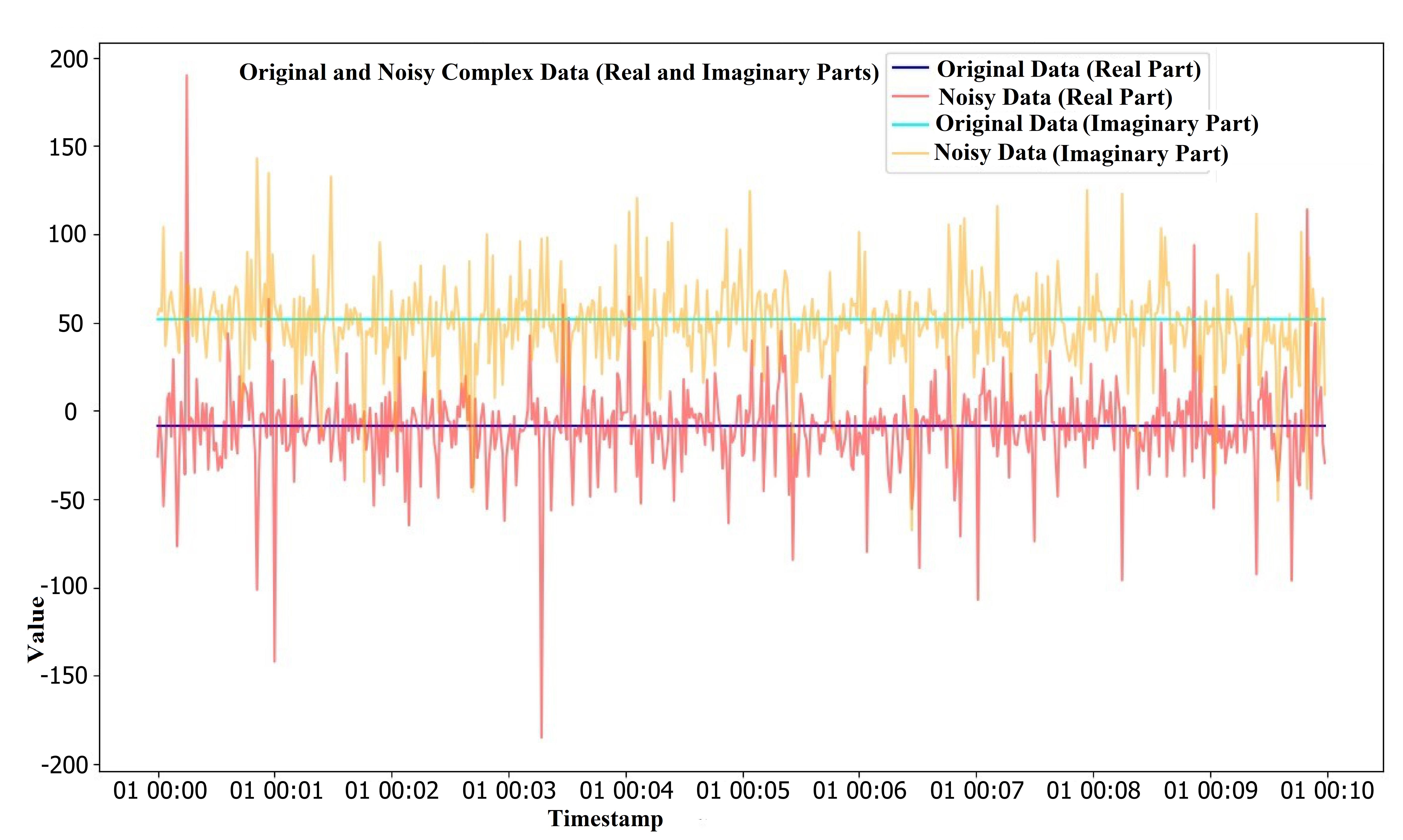} 
    \caption{A rolling window approach with a 2-minute window size was applied to the 10-minute dataset.}
\end{figure*}

\par
Fig. 5 demonstrates how a 3000-data-point sample was carefully chosen to train the LSTM model for computational efficiency and effectiveness. This method focuses on the dataset's main properties without overfitting from huge datasets. Training on this representative sample improved generalization and channel predictions by teaching the model important data patterns and relationships.
\begin{figure}[htbp]
    \centering
    \includegraphics[width=\columnwidth]{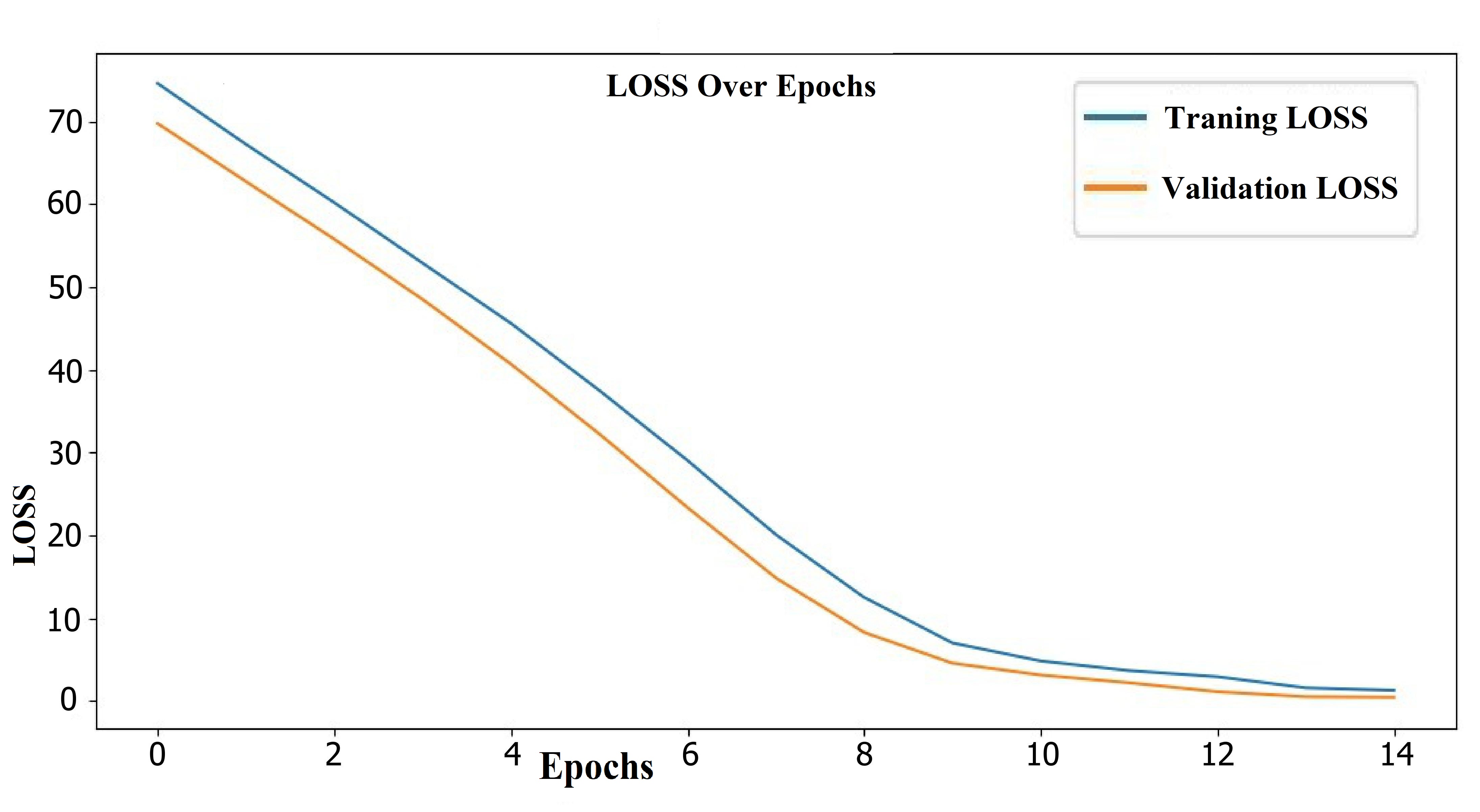} 
    \caption{Loss Function of Training and Validation Dataset.}
\end{figure}
\par
Received pilot signals, power allocation vectors, and partly decoded data symbols improve NOMA channel estimate accuracy compared to a CPF that only uses received pilot signals. Further information about the subcarrier's fractional power allotment helps the CPF capture NOMA systems' dynamic channel fluctuations and interference. This revised CPF improves SNR performance over traditional pilot-based estimating approaches (Fig. 6).
\begin{figure}[htbp]
    \centering
    \includegraphics[width=\columnwidth]{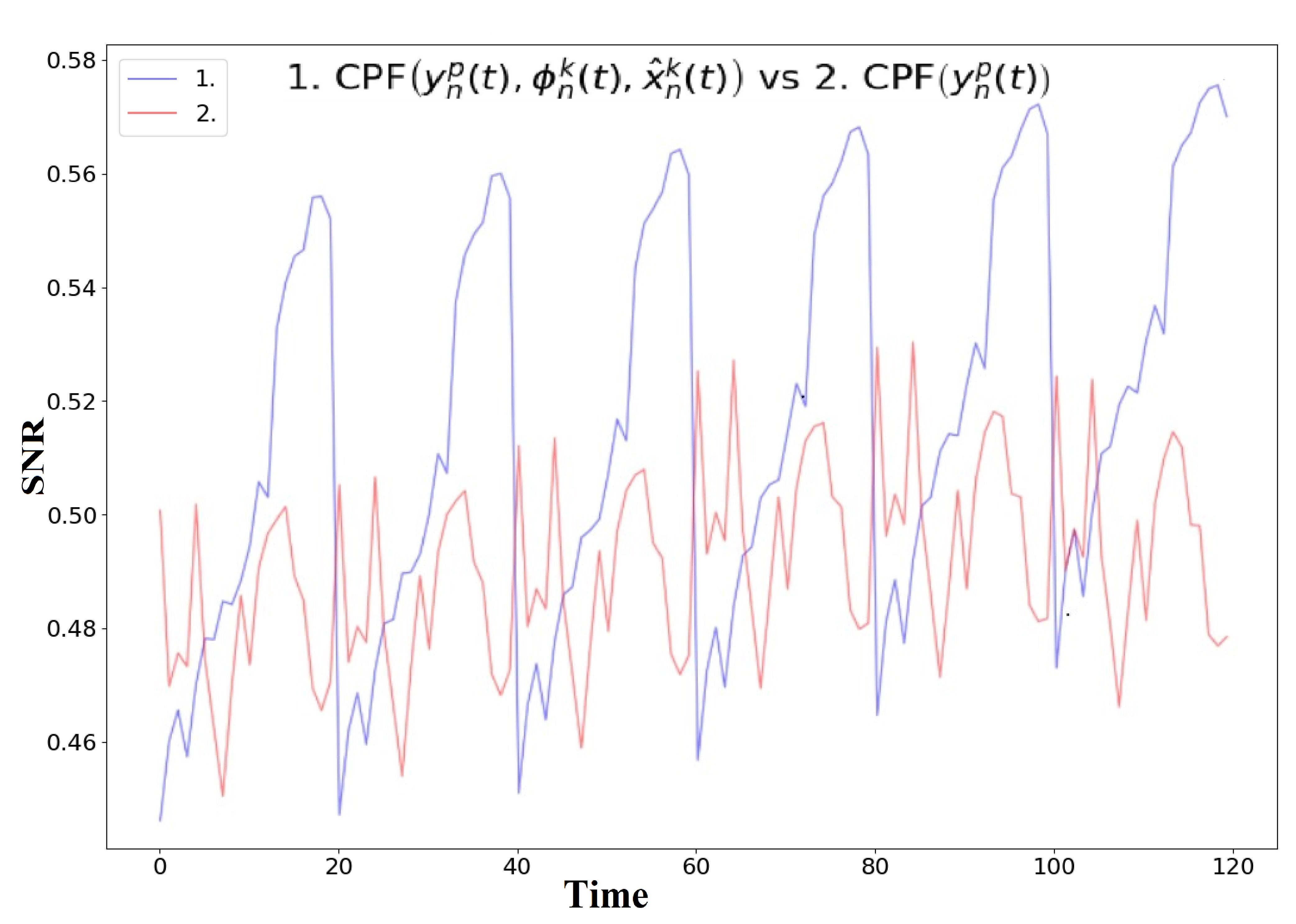} 
    \caption{Enhanced Channel Estimation in NOMA via CPF with Fractional Power Allocation.}
\end{figure}
\subsection{Addressing Scalability Challenges for Larger Networks}
\begin{figure}[htbp]
    \centering
    \includegraphics[width=\columnwidth]{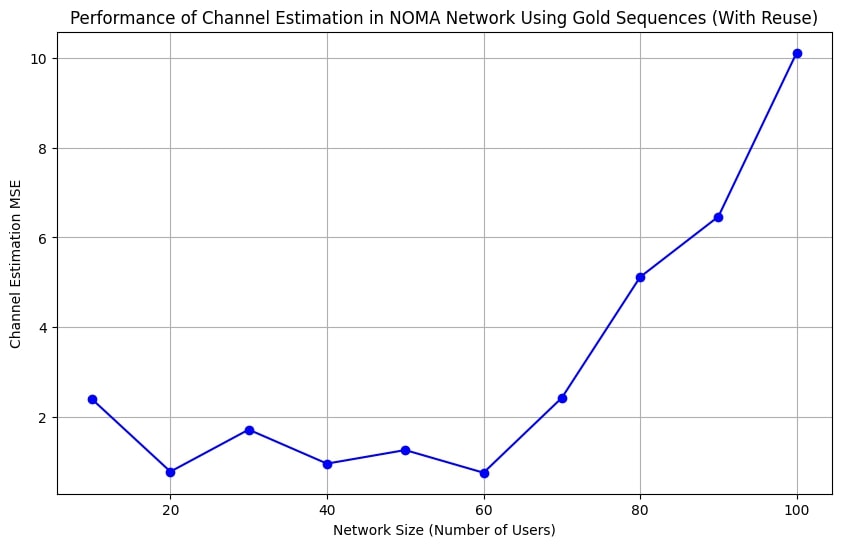} 
    \caption{Channel Estimation in NOMA Networks Using Gold Sequences: Evaluating Scalability, Interference, and MSE.}
    \label{fig:chap_6.2}
\end{figure}

In Fig~\ref{fig:chap_6.2}, simulates and evaluates the performance of channel estimation in the NOMA network using Gold sequences, which are ideal for multi-user environments due to their excellent correlation properties and low cross-correlation. The process begins with generating Gold sequences by combining two preferred pairs of maximum-length sequences (m-sequences) using linear feedback shift registers (LFSRs) with specific polynomial coefficients. The simulation models a NOMA network with varying numbers of users, incorporating sequence reuse to accommodate larger networks, which introduces interference, a realistic challenge in large-scale systems. Channel estimation is performed using matched filtering, where received signals are correlated with the transmitted Gold sequences to estimate channel coefficients. The performance is evaluated using the Mean Squared Error (MSE) metric, with results showing that as for the larger network of size 40 to 100 UEs sequences are reused, and interference grows, leading to higher MSE values. Despite this, Gold sequences maintain relatively low MSE values, demonstrating their robustness in multi-user environments. However, the simulation reveals scalability limitations, as performance degrades in very large networks (> 60 UEs) due to increased interference. The findings emphasize the importance of sequence design and interference management in NOMA systems, highlighting the need for advanced techniques like interference cancellation or adaptive sequence allocation to improve scalability. Limitations include the use of a fixed Gold sequence length (31), a simplified interference model, suggesting future work should explore adaptive sequence lengths, more realistic interference models.

Gold sequences are widely used in communication systems, including NOMA, due to their good correlation properties and low cross-correlation values, making them suitable for user separation and interference mitigation in multi-user environments. However, their scalability for larger networks in NOMA systems depends on several factors. First, Gold sequences are generated using preferred pairs of m-sequences with a length of $2^n - 1$, where $n$ is the degree of the polynomial used to generate the m-sequences. As the network size grows, the number of required sequences increases, and if the number of users exceeds the available Gold sequences, sequences must be reused, leading to increased interference. To support larger networks, the sequence length must be increased, which raises the complexity of sequence generation and processing. Second, Gold sequences have bounded cross-correlation values, which are crucial for minimizing interference between users, but as the network size grows, cross-correlation may become a limiting factor, especially with sequence reuse, degrading system performance through increased multi-user interference (MUI). Third, generating and processing longer Gold sequences requires more computational resources, posing challenges for larger networks, particularly in real-time systems, while the complexity of detecting and decoding signals in NOMA systems also increases with the number of users and sequence length. Fourth, NOMA relies on superposition coding and successive interference cancellation (SIC) to achieve high spectral efficiency, and Gold sequences must be designed to ensure that the SIC process is not overly complicated by high cross-correlation or interference, with the trade-off between sequence length, correlation properties, and spectral efficiency becoming critical for larger networks. Fifth, if Gold sequences are not scalable for very large networks, alternative sequence designs or multiple access techniques, such as Zadoff-Chu sequences, random sequences, or hybrid approaches combining Gold sequences with power-domain or spatial-domain NOMA, may be considered. Finally, in practice, the scalability of Gold sequences for NOMA systems depends on the specific application, channel conditions, and hardware limitations, with simulation and testing essential to evaluate their performance in larger networks and determine whether they meet the required quality of service (QoS) metrics. In conclusion, Gold sequences are effective for small to medium-sized NOMA networks due to their excellent correlation properties, but for larger networks, their scalability is limited by sequence length, cross-correlation, and computational complexity, necessitating alternative sequences or hybrid approaches to ensure efficient and reliable communication in large-scale NOMA systems.
\section{Conclusion}
Gold coding NOMA system performance is compared to C-V-BLAST methods in this study. The findings show that gold-coded NOMA has better SER across SNR levels. Due to their variety, Gold sequences help overcome spatial constraints. A dataset-trained deep learning model network predicted channel activity accurately. The proposed channel estimation approach uses received pilot signals, power allocation vectors, and data symbols to outperform pilot-based methods. Several fields might use the presented technique and results. Identifying how noise components and channel conditions affect system performance may help network operators optimize resource allocation, power control, and interference management.

\bibliographystyle{plain}
\bibliography{sample-base}
\end{document}